# Possible magnetism based on orbital motion of protons in ice

Fei Yen[1,*], Tian Gao[2], Yongsheng Liu[2], Adam Berlie[3,4]

[1]Key Laboratory for Materials Physics, Institute of Solid State Physics, Hefei Institutes of Physical Science, Chinese Academy of Sciences, Hefei 230031, P. R. China

[2]School of Mathematics and Physics, Shanghai University of Electric Power, No. 2588 Changyang Road, Shanghai 200090, P. R. China

[3]ISIS Neutron and Muon Source, STFC Rutherford Appleton Laboratory, Harwell Science and Innovation Campus, Didcot, Oxfordshire OX11 0QX, United Kingdom

[4]RIKEN Nishina Center for Accelerator-Based Science, 2-1 Hirosawa, Wako, Saitama 351-0198, Japan.

*Correspondence: fyen18@hotmail.com, fyen@issp.ac.cn,

**Abstract:** A peak anomaly is observed in the magnetic susceptibility as a function of temperature in solid $H_2O$ near $T_p$=60 K. At external magnetic fields below 2 kOe, $T_p$ becomes positive in the temperature range between 45 and 66 K. The magnetic field dependence of the susceptibility in the same temperature range exhibits an inverted ferromagnetic hysteretic loop superimposed on top of the diamagnetic signature of ice at fields below 600 Oe. We suggest that a fraction of protons that are capable of undergoing correlated tunneling in a hexagonal path without disrupting the stoichiometry of the lattice create an induced magnetic field opposite to the induced magnetic field created by the electrons upon application of an external field which counters the overall diamagnetism of the material.

Most ice in nature is designated as ice I*h* and it is a unique system in that there is long range ordering of the oxygen atoms in a hexagonal pattern while the hydrogen atoms are distributed in a disordered fashion throughout the lattice [1]. Only two rules must be obeyed [2]: #1 that only one hydrogen atom can reside in between each pair of oxygen atoms and #2 each oxygen atom must be attached to two hydrogen atoms via a covalent bond (about 1.00 A in length [3]) and to another two hydrogen atoms via a hydrogen bond (about 1.75 A in length). The second rule allows the hydrogen atoms to rest in one of two available sites between each pair of oxygen atoms. All of the electrons form a band insulator so the hydrogen atoms are reduced to protons [4]. From such, hexagonal ice is a superposition of two subsystems: crystalline oxygen atoms and glassy protons (Fig. 1).

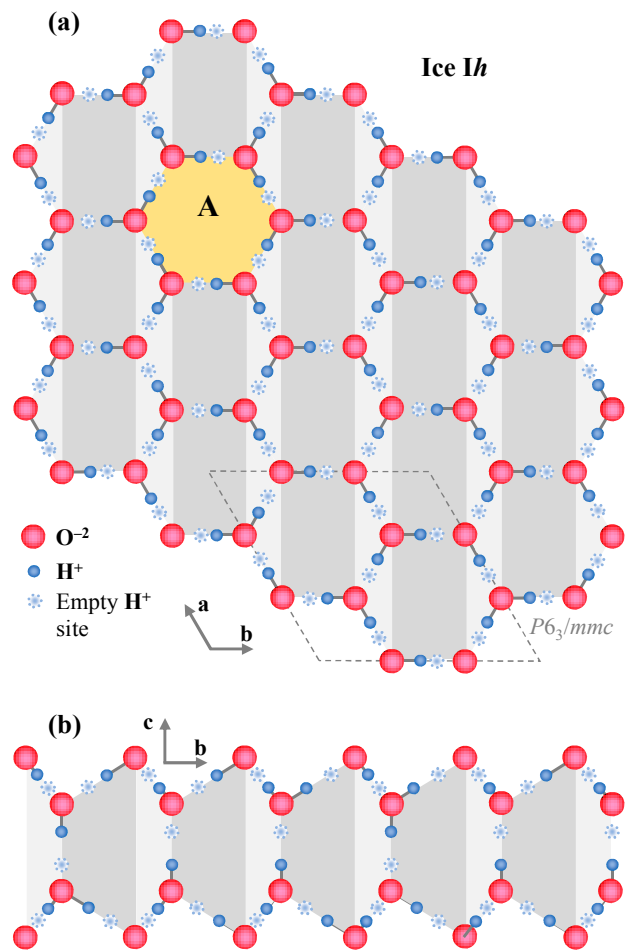

**Fig. 1:** (a) Basal plane of hexagonal ice I*h*. The oxygen atoms form hexagonal staircase layers along the c-axis direction. Between each pair of oxygen atoms one proton occupies one out of two available sites in a quasi-random fashion as they still have to obey the ice rules. The hexamer labeled A is 1 in 36 that has all of its protons aligned in a clockwise direction. Dashed diamond is the unit cell. (b) The prismatic face of hexagonal ice I*h*.

Figure 1a shows the basal plane of ice I*h* which consists of highly symmetric hexamers with oxygen atoms residing at the vertices with one proton on each side occupying one out of their respective two available sites. The basal plane has a staircase type of pattern which is more evident if viewed from the side as shown in Fig. 2b. Ice also has a prismatic face parallel to its c-axis. At around 72 K, the disordered protons are expected to become ordered [5]. However, proton hopping by classical means cannot take place at such low a temperature from its glass transition of $T_g$=136 K [6]. Quantum tunneling of a proton to its adjacent site has been suggested to possibly take place since the wave function of the proton stretches all the way to the oxygen atoms at both ends [7]. However, tunneling of one proton alone would create an $OH^-$ and $H_3O^+$ defect which is energetically unfavorable. As such, the protons in ice remain mostly disordered down to the lowest temperatures which is why there exists a residual entropy at absolute zero [8]. At even lower temperatures, when the thermal fluctuations are low enough, the tendency of the protons to want to become ordered (lower its energy) causes them to tunnel in concert to their respective sites. The hexamer labeled A in Fig. 1a shows how for the particular configuration when all of the protons are aligned clockwise or anticlockwise, that a concerted tunneling of all of these six protons to their adjacent empty sites is possible without violating any of the two ice rules. Quantum simulations and calculations have shown that this is not only possible but most probable [9,10,11]. Quasielastic neutron scattering have also shown that highly correlated mobility of protons in groups occur at 5 K [12]. Dielectric constant measurements have also suggested that once the protons tunnel in concert to their adjacent sites, that the protons tunnel back to their original sites because the system remains disordered [13]. The continued tunneling between their respective available sites renders the protons two stationary states. Upon the tunneling of six protons whether clockwise or anticlockwise, the movement of charge about an almost enclosed area should yield a magnetic moment. It is therefore interesting to know if there are any exotic magnetic properties accompanying this new quantum phenomenon; especially when the diamagnetic properties of ice have not been thoroughly studied at temperatures below 77 K due to the notion that the diamagnetic signal should be independent of temperature.

In this Letter, we report on a previously unidentified peak anomaly in the magnetization of ice occurring near 60 K up to applied external magnetic fields of 50 kOe. The magnetization even becomes positive in the temperature range of 45-66 K at external magnetic fields lower than 2 kOe. In the field versus magnetization isotherms between 45 and 66 K, an inverted ferromagnetic hysteresis loop is observed at fields lower than ±600 Oe. We discuss and suggest how most likely, correlated intermolecular tunneling of protons in hexagonal loops, which mimic a set of positive charges in orbital motion, generate an induced magnetic field that partially counters the induced magnetic field created by the negative charges of the electrons.

Liquid $H_2O$ with a resistivity of 18.2 MΩ-cm from a Milli-Q Direct 8 dispenser was poured into a Teflon container, topped off, sealed with a stainless steel cap, and mounted onto the measurement probe of the VSM option of a Physical Properties Measurement System (PPMS) manufactured by Quantum Design. The measurement amplitude was set to 1 mm with the factory frequency of 40 Hz and average time of 5

s. Centering was set at 0.5 mm above the bottom of the Teflon container so the signal contained a weak but constant paramagnetic background. The standard deviation of the magnetization was always about two orders or magnitude lower than the magnetization at $10^{-7}$ emu. The sample was cooled at 2 K/min where hexagonal ice usually forms from supercooled water near 255 K according to previous dielectric constant measurements [14]. For the field dependent measurements, the applied fields were ramped by no more than 200 Oe/s. No traces of oxygen or nitrogen impurities were detected from dielectric constant measurements which are extremely sensitive to even solid-solid phase transitions of impurities. For the $D_2O$ experiments, 99.9% purity, 1.107 g/mL liquid $D_2O$ from Adelma Reagents Co., Ltd. was used as the sample.

Figure 2a shows the temperature dependence of the magnetization $M(T)$ at 2 and 10 kOe with three different samples for $H_2O$. Fig. 2b shows $M(T)$ for $D_2O$ up to 30 kOe. The data above 100 K for fields above 10 kOe are nearly temperature independent and coincides well with previous reports [15]. In all curves, a peak is observed near 60 and 62 K for $H_2O$ and $D_2O$, respectively. The peak near 60 K is most likely a consequence of the expansion coefficient of ice becoming negative at the same temperature [16]. The dielectric constant of ice also exhibits a discontinuity anomaly near 60 K and 62 K for $H_2O$ and $D_2O$, respectively [17].

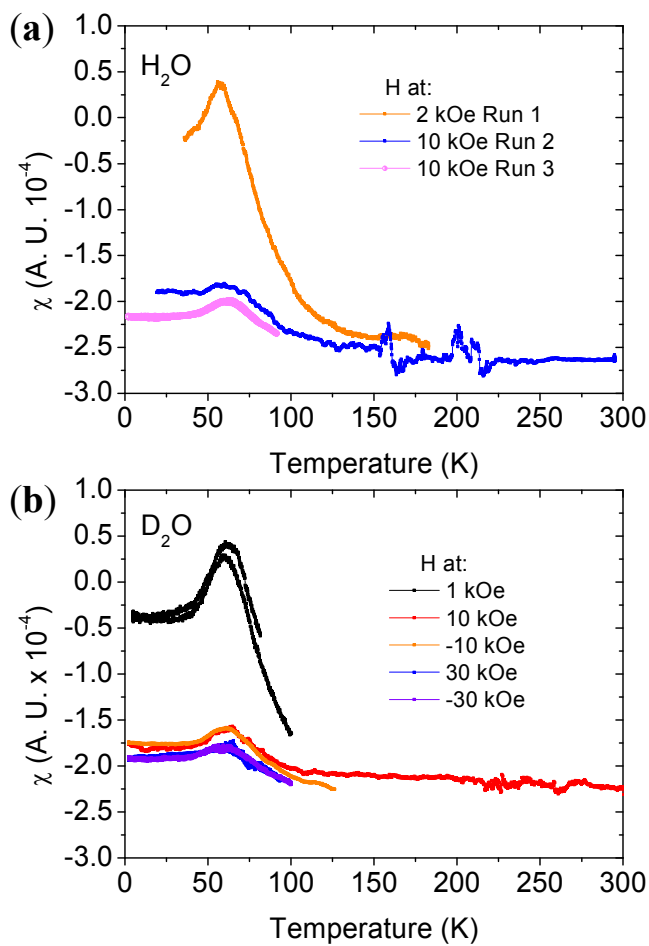

**Fig. 2:** Magnetization as a function of temperature $M(T)$ curves of $H_2O$ and $D_2O$.

Figure 3a displays the magnetization as a function of external magnetic field $M(H)$ at various temperatures in the field range of $12 < H < 22$ kOe. For the isotherms below 36 K and above 75 K, the magnetic susceptibility is nearly independent up to 50 kOe. In contrast, the 59 K isotherm exhibits a magnitude that is 9% smaller in the field range of $3 > H > 50$ kOe which coincides with the peak anomaly in $M(T)$. Figs. 3b - 3f show an enlarged view of the low field region of isotherms at 45, 54, 59, 66 and 75 K. The most unusual and apparent feature in the 45-66 K isotherms is the inverted hysteretic loop centered at the origin. For the case of the 59 K isotherm, the region extends from about –600 to 600 Oe. In this temperature range, the magnetization does not intersect the origin if the data at $H >1$ kOe is extrapolated to zero as it is supposed to for a diamagnetic system. In contrast, the magnetization at low fields obtained for the 75 K isotherm appears completely normal. Fig. 3g shows an enlarged view of the 59 K isotherm with the addition of two more curves that were measured at different initial conditions. Curves 1 and 2 were measured during decreasing and increasing $H$, respectively, from +50 to –50 kOe and vice versa. Curve 3 is the measured magnetization while $H$ was decreased at a rate of 5 Oe/s from +200 to –200 Oe *after* $H$ was ramped from –2 kOe to +200 Oe at 100 Oe/s. It is evident that Curve 3 tends to return back to its original Curve 2 rather than crossing over to the opposing Curve 3 such as in ferromagnetic systems. The history of Curve 4 is similar to Curve 3, but instead of stopping at +200 Oe, $H$ was set to +500 Oe. This time, the magnetization underwent a different route, but nevertheless, it tends to trace back to its original path. This effect is noticeable up to magnetic fields near 2 kOe; with higher magnetic fields set as their initial points, the magnetization traces either Curves 1 or 2 depending on the polarity of the field. Data with odd symmetry to Curves 1 and 2 were observed if the same history, but with opposite polarity, was employed. If attention was only paid to Fig. 3a, one would presume that the magnetization values of the $M(T)$ curves at low temperatures ($T<36$ K) should be near the same. However, from Fig. 3g, it is evident that the magnetization at low temperatures depends on the history of $H$, and especially when $H$ is lower than 2 kOe.

The electrons in ice are rigidly held in their orbitals. The oxygen atoms are also non-magnetic so the peak anomaly near 60 K and inverted hysteresis loop are not consequences of paramagnetism and ferromagnetism, respectively. We also rule out the possibility of a magnetic contribution from impurity phases as the paramagnetic and ferromagnetic magnetizations should increase with decreasing temperature. We note that the absolute values of the magnetization at ±50 kOe for the isotherms of 5, 12, 19, 27, 36 and 75 K shown in Fig. 3a are all the same.

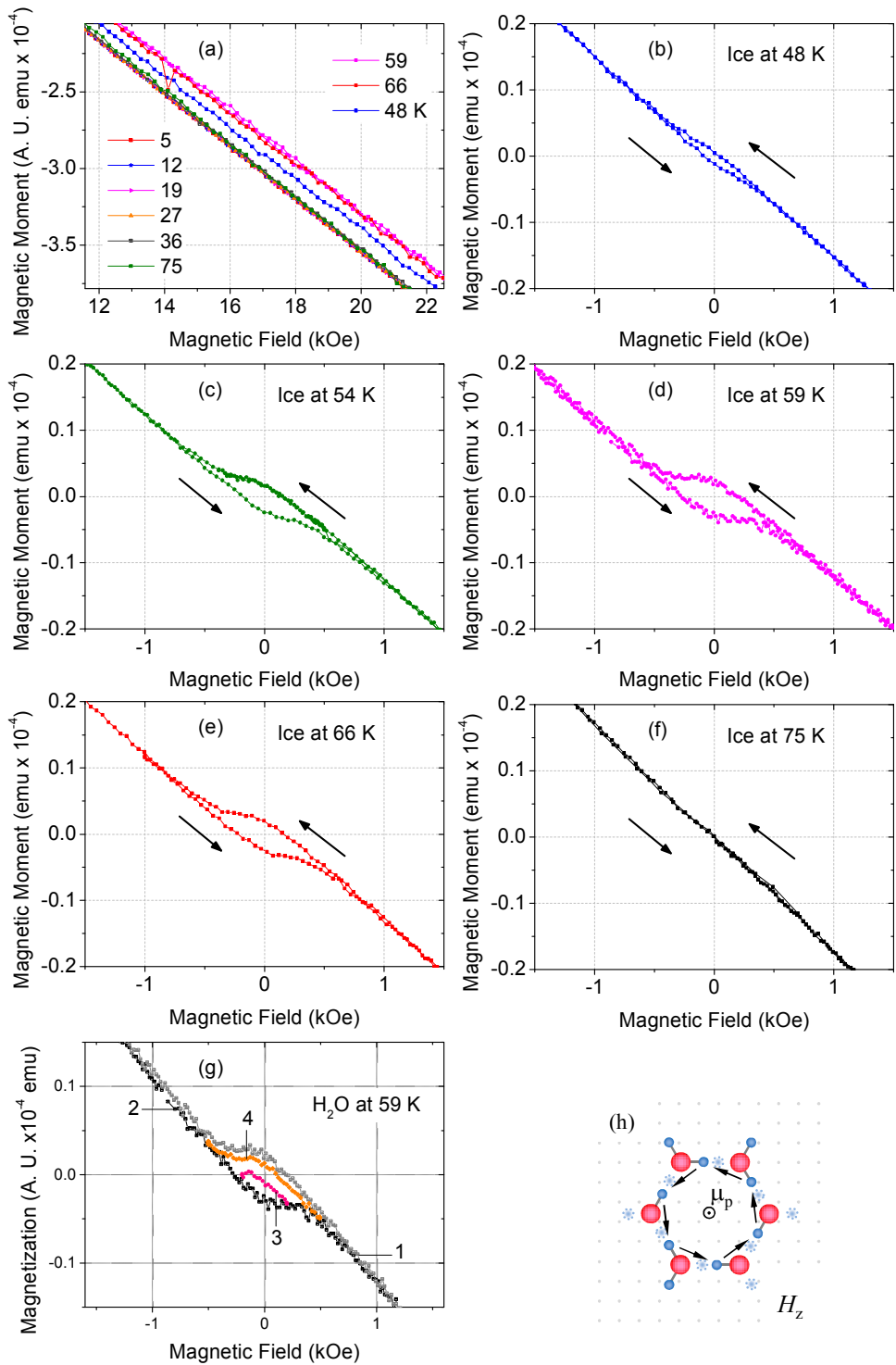

**Fig. 3: (a)** Magnetization as a function of external magnetic field $M(H)$ curves of $H_2O$ at various temperatures. **(b)-(f)** show hysteretic loops at low fields. (g) shows four curves subjected to different initial conditions Curve 1 measured from 50 kOe → –50 kOe; Curve 2: –50 kOe → 50 kOe; Curve 3: –2 kOe → 200 Oe → –200 Oe; and Curve 4: –2 kOe → 500 Oe → –500 Oe. **(h)** portrays how an applied field $H_z$ likely induces a concerted proton tunneling of six protons (depicted by the arrows) which create an associated magnetic moment $\mu_p$ to point along the same direction of $H_z$.

It is a well established fact that upon genetic mutation of a cell, the elemental process taking place is a two-body proton tunneling between two nucleotides in DNA [18,19], and the probability of tunneling increases if an external source of energy is supplied such as radiation. Simultaneous two-body proton tunneling is also believed to take place in carboxylic acid dimers [20] and shown that due to the competition between synchronous and non-synchronous tunneling processes, the trajectories taken do not necessarily occur through the narrowest potential barriers [21]. Isaacs *et al.*, employed a quantum mechanical model which takes into account phase coherence of neighboring molecules to describe their experimental results obtained on the Compton profile for ice I$h$ [3]. This showed the high degree of covalency between neighboring molecules via the hydrogen bonds. It is therefore not far-fetched to imagine that with the aid of an external magnetic field that the protons overcome their respective potential barriers to undergo intermolecular tunneling. Tunneling of one proton, or two, or even up to five-body tunneling results in a violation of the ice rules. However, for a particular hexamer with the type of proton configuration shown in Fig. 3h, if the six protons tunneled in concert as shown by the arrows, the ice rules are preserved [22]. Successive jumps of the protons encircling around a hexagonal orbit at a tunneling rate of $\tau$ induces a current $I=q\tau$, where $q$ is the elemental unit of charge. The product of the current with the area of the hexamer gives rise to a local magnetic moment $\mu_p$ which is opposite in direction to the individual magnetic moments created by the electrons $\mu_e$. The net effect is an overall reduction of the magnetization of the system which qualitatively explains our experimental results.

To our knowledge, there is no existing literature on magnetism based on orbital motion of protons in a solid. According to Langevin's theory of diamagnetism [23], the magnetic moment of an electron is

$$\mu_e = -\frac{Zq^2 H}{4m_e} r_e^2 \qquad (1)$$

where $Z$ is the atomic number, $m_e$ is the mass of the electron and $r_e$ the average radius of the electrons orbiting perpendicular to $H$. With slight alterations, the induced magnetic moment perpendicular to the basal plane in hexagonal ice contributed by six protons can be estimated to be:

$$\mu_p = 6\tau(H) \frac{q^2 H}{4m_p} r_p^2 \qquad (2)$$

where $m_p$ and $r_p$ are the mass and approximate distance of the protons from the center of the hexamer. $m_p$ is around 200 times that of $m_e$, however, the ratio of $r_p/r_e$ far exceeds 200. Hence, $\mu_p$ should be larger than $\mu_e$. The most important feature in Eq. 2 is that $\tau(H)$ is a function of $H$ and should be proportional to $H$. This means that with increasing $H$, $\tau$ should also increase accordingly. However, $\tau$ must have an upper limit because each time the protons tunnel, they must first stabilize before being able to tunnel again which requires a time delay. At high values of $H$, the protons can only tunnel so fast unlike the electrons which can accelerate to speeds nearing the speed of light. Hence, at low $H$, i.e. at $H$<200 Oe at 59 K, the overall magnetization $M$ is positive because the magnetization contribution of the protons [24] $M_p$ is larger than that of the electrons $M_e$. However, at higher fields, the overall magnetization $M$ becomes negative because $M_p$ is saturated while $M_e$ increases linearly with respect to $H$. This explains the inverted ferromagnetic loops observed in Figs. 3b-3e. According to Figs. 3e and 3f, it is apparent that $M$ becomes positive somewhere in between 66 and 75 K which coincides with the phase transition between the proton ordered (ice XI) and proton disordered phases (ice I$h$) at $T_c$=72 K [17]. If $H$ is zero while cooling through $T_c$, $M$ should also be zero, because all of the individual $\mu_p$ moments are randomly oriented which cancels out any net effect, akin to randomly oriented ferromagnetic domains possessing a near zero $M$. In contrast, at small values of $H$, $M$ should become positive near $T_c$ and the values of $M$ at lower temperature may depend on the history which explains the results we obtained on our $M(T)$ experiments. Ice I$h$ possesses the smallest volume at 60 K [16], at temperatures below it starts to expand again. Naturally, the intermolecular distances should also exhibit a minimum at this temperature. Since the probability of quantum tunneling depends highly on the width of its potential barrier, it is therefore expected that a maximum also exists in $M$ near 60 K, which coincides with the peak anomaly observed in our results. Interestingly, annealing doped samples of ice I$h$ in the range of 57-63 K for several days often yields the highest concentrations of ice XI [25,26]. It seems that the thermodynamic phase transition from ice I$h$ to ice XI should also occur near 72 K, but it does not take place until near 60 K because the only allowed type of proton mobility at such temperatures appears to be through the correlated tunneling process as suggested in Fig. 3h.

Apart from the ordering of nuclear spins at ultra-low temperatures [27], most of the magnetism in solids is due to the way electrons behave. Magnetism based on the motion of protons is an entirely new phenomenon. A more quantitative approach is definitely needed to obtain the shape of the potential barriers so that the tunneling rates, tunneling times and excitation energies can be determined. The proposed magnetic moment contribution stemming from the protons (Eq. 2) is also a rough approximation as we do not take into consideration any possible effects arising from spin-orbit & spin-lattice coupling, quantum nuclear effects, competing quantum effects [28], and the internal electric field. The highly correlated orbital motion of the protons is also likely to be mediated by phonons such as Cooper pairs in superconductors. From such, the correlated cyclic motion of protons in groups of six may even comprise a new type of quasi-particle. We propose quasi-elastic neutron scattering and Compton profile anisotropy experiments on ice near 60 K under magnetic fields of 1-2 kOe to check if jumping distances of around 2.2 Å exist to further verify this phenomenon.

**Acknowledgements:**

This work was made possible by the National Natural Science Foundation of China, grant numbers 11374307, 11250110050, 11204171 and 11374204.

configurations that are disordered, a concerted tunneling of all of its six protons will result in a violation of the ice rules so only 1 in 18 hexamers may possess a magnetic moment associated to their orbital motion.